\documentstyle[twoside,epsf]{article}

\catcode`\@=11
\long\def\@makefntext#1{
\protect\noindent \hbox to 3.2pt {\hskip-.9pt  
$^{{\eightrm\@thefnmark}}$\hfil}#1\hfill}               

\def\@makefnmark{\hbox to 0pt{$^{\@thefnmark}$\hss}}    
        
\def\ps@myheadings{\let\@mkboth\@gobbletwo
\def\@oddhead{\hbox{}
\rightmark\hfil\eightrm\thepage}   
\def\@oddfoot{}\def\@evenhead{\eightrm\thepage\hfil
\leftmark\hbox{}}\def\@evenfoot{}
\def\sectionmark##1{}\def\subsectionmark##1{}}

\oddsidemargin=\evensidemargin
\newcounter{sectionc}\newcounter{subsectionc}\newcounter{subsubsectionc}
\renewcommand{\section}[1] {\vspace{12pt}\addtocounter{sectionc}{1} 
\setcounter{subsectionc}{0}\setcounter{subsubsectionc}{0}\noindent 
        {\tenbf\thesectionc. #1}\par\vspace{5pt}}
\renewcommand{\subsection}[1] {\vspace{12pt}\addtocounter{subsectionc}{1} 
        \setcounter{subsubsectionc}{0}\noindent 
        {\bf\thesectionc.\thesubsectionc. {\kern1pt \bfit #1}}\par\vspace{5pt}}
\renewcommand{\subsubsection}[1] {\vspace{12pt}\addtocounter{subsubsectionc}{1}
        \noindent{\tenrm\thesectionc.\thesubsectionc.\thesubsubsectionc.
        {\kern1pt \tenit #1}}\par\vspace{5pt}}
\newcommand{\nonumsection}[1] {\vspace{12pt}\noindent{\tenbf #1}
        \par\vspace{5pt}}

\newcounter{appendixc}
\newcounter{subappendixc}[appendixc]
\newcounter{subsubappendixc}[subappendixc]
\renewcommand{\thesubappendixc}{\Alph{appendixc}.\arabic{subappendixc}}
\renewcommand{\thesubsubappendixc}
        {\Alph{appendixc}.\arabic{subappendixc}.\arabic{subsubappendixc}}

\renewcommand{\appendix}[1] {\vspace{12pt}
        \refstepcounter{appendixc}
        \setcounter{figure}{0}
        \setcounter{table}{0}
        \setcounter{lemma}{0}
        \setcounter{theorem}{0}
        \setcounter{corollary}{0}
        \setcounter{definition}{0}
        \setcounter{equation}{0}
        \renewcommand{\thefigure}{\Alph{appendixc}.\arabic{figure}}
        \renewcommand{\thetable}{\Alph{appendixc}.\arabic{table}}
        \renewcommand{\theappendixc}{\Alph{appendixc}}
        \renewcommand{\thelemma}{\Alph{appendixc}.\arabic{lemma}}
        \renewcommand{\thetheorem}{\Alph{appendixc}.\arabic{theorem}}
        \renewcommand{\thedefinition}{\Alph{appendixc}.\arabic{definition}}
        \renewcommand{\thecorollary}{\Alph{appendixc}.\arabic{corollary}}
        \renewcommand{\theequation}{\Alph{appendixc}.\arabic{equation}}
        \noindent{\tenbf Appendix \theappendixc #1}\par\vspace{5pt}}
\newcommand{\subappendix}[1] {\vspace{12pt}
        \refstepcounter{subappendixc}
        \noindent{\bf Appendix \thesubappendixc. {\kern1pt \bfit #1}}
        \par\vspace{5pt}}
\newcommand{\subsubappendix}[1] {\vspace{12pt}
        \refstepcounter{subsubappendixc}
        \noindent{\rm Appendix \thesubsubappendixc. {\kern1pt \tenit #1}}
        \par\vspace{5pt}}

\newcommand{\textlineskip}{\baselineskip=13pt}
\newcommand{\smalllineskip}{\baselineskip=10pt}

\def\eightcirc{
\begin{picture}(0,0)
\put(4.4,1.8){\circle{6.5}}
\end{picture}}
\def\eightcopyright{\eightcirc\kern2.7pt\hbox{\eightrm c}} 

\newcommand{\copyrightheading}[1]
        {\vspace*{-2.5cm}\smalllineskip{\flushleft
        {\footnotesize International Journal of Modern Physics C, #1}\\
        {\footnotesize $\eightcopyright$\,\,\, World Scientific Publishing
         Company}\\
         }}

\newcommand{\publisher}[2]{{\begin{center}\footnotesize\smalllineskip 
        Received #1\\
        Revised #2
        \end{center}
        }}

\def\abstracts#1#2#3{{
        \centering{\begin{minipage}{4.5in}\baselineskip=10pt\footnotesize
        \parindent=0pt #1\par
        \parindent=15pt #2\par
        \parindent=15pt #3\par
        \end{minipage}}\par}} 

\def\keywords#1{{
       \centering{\begin{minipage}{4.5in}\baselineskip=10pt\footnotesize
       {\footnotesize\it Keywords}\/: #1
        \end{minipage}}\par}}

\newcommand{\bibit}{\nineit}
\newcommand{\bibbf}{\ninebf}
\renewenvironment{thebibliography}[1]
        {\frenchspacing
         \ninerm\baselineskip=11pt
         \begin{list}{\arabic{enumi}.}
        {\usecounter{enumi}\setlength{\parsep}{0pt}     
         \setlength{\leftmargin 17pt}{\rightmargin 0pt}   
         \setlength{\itemsep}{0pt} \settowidth
        {\labelwidth}{#1.}\sloppy}}{\end{list}}

\newcounter{itemlistc}
\newcounter{romanlistc}
\newcounter{alphlistc}
\newcounter{arabiclistc}

\newcommand{\fcaption}[1]{
        \refstepcounter{figure}
        \setbox\@tempboxa = \hbox{\footnotesize Fig.~\thefigure. #1}
        \ifdim \wd\@tempboxa > 5in
           {\begin{center}
        \parbox{5in}{\footnotesize\smalllineskip Fig.~\thefigure. #1}
            \end{center}}
        \else
             {\begin{center}
             {\footnotesize Fig.~\thefigure. #1}
              \end{center}}
        \fi}

\newcommand{\tcaption}[1]{
        \refstepcounter{table}
        \setbox\@tempboxa = \hbox{\footnotesize Table~\thetable. #1}
        \ifdim \wd\@tempboxa > 5in
           {\begin{center}
         \parbox{5in}{\footnotesize\smalllineskip Table~\thetable. #1}
            \end{center}}
        \else
             {\begin{center}
             {\footnotesize Table~\thetable. #1}
              \end{center}}
        \fi}

\def\@citex[#1]#2{\if@filesw\immediate\write\@auxout
        {\string\citation{#2}}\fi
\def\@citea{}\@cite{\@for\@citeb:=#2\do
        {\@citea\def\@citea{,}\@ifundefined
        {b@\@citeb}{{\bf ?}\@warning
        {Citation `\@citeb' on page \thepage \space undefined}}
        {\csname b@\@citeb\endcsname}}}{#1}}

\newif\if@cghi
\def\cite{\@cghitrue\@ifnextchar [{\@tempswatrue
        \@citex}{\@tempswafalse\@citex[]}}
\def\citelow{\@cghifalse\@ifnextchar [{\@tempswatrue
        \@citex}{\@tempswafalse\@citex[]}}
\def\@cite#1#2{{$\null^{#1}$\if@tempswa\typeout
        {IJCGA warning: optional citation argument 
        ignored: `#2'} \fi}}

\def\pmb#1{\setbox0=\hbox{#1}
        \kern-.025em\copy0\kern-\wd0
        \kern.05em\copy0\kern-\wd0
        \kern-.025em\raise.0433em\box0}

\def\fnt#1#2{\footnotetext{\kern-.3em
        {$^{\mbox{\scriptsize #1}}$}{#2}}}

\def\fpage#1{\begingroup
\voffset=.3in
\thispagestyle{empty}\begin{table}[b]\centerline{\footnotesize #1}
        \end{table}\endgroup}

\def\runninghead#1#2{\pagestyle{myheadings}
\markboth{{\protect\footnotesize\it{\quad #1}}\hfill}
{\hfill{\protect\footnotesize\it{#2\quad}}}}
\font\tenbf=cmbx10
\font\tenit=cmti10 
\font\tenit=cmti10
\font\bfit=cmbxti10 at 10pt
\font\ninebf=cmbx9
\font\ninerm=cmr9
\font\nineit=cmti9

\font\eightrm=cmr8

\def\lsym{\raise-3pt\hbox{\vbox{\tabskip0pt\offinterlineskip
        \halign{\tabskip0pt plus 1em
        ##\tabskip0pt\cr
        $\,\,<\,\,$\cr
        $\,\,\sim\,\,$\cr}}}}
\def\rsym{\raise-3pt\hbox{\vbox{\tabskip0pt\offinterlineskip
     \halign{\tabskip0pt plus 1em
      ##\tabskip0pt\cr
      $\,\,>\,\,$\cr
      $\,\,\sim\,\,$\cr}}}}
\def\qed{\hbox{${\vcenter{\vbox{                        
        \hrule height 0.4pt\hbox{\vrule width 0.4pt height 6pt
        \kern5pt\vrule width 0.4pt}\hrule height 0.4pt}}}$}}
\def\theequation{\thesection.\arabic{equation}}         

\def\nle{\ \raise.3ex\hbox{$<$}\kern-0.8em\lower.7ex\hbox{$\sim$}\ }
\def\nge{\ \raise.3ex\hbox{$>$}\kern-0.8em\lower.7ex\hbox{$\sim$}\ }

\def\DelCone{1 - C_T(\tau; t_{\rm w})}
\def\DelConetwo{[1 - C_{T_2,T_1}(\tau; t_{\rm w1})]}
\def\DelC{\Delta C_T(\tau; t_{\rm w})}
\def\DelCtwo{\Delta C_{T_2,T_1}(\tau; t_{\rm w1})}

\def\Tc{T_{\rm c}}
\def\tw{t_{\rm w}}
\def\twone{t_{\rm w1}}
\def\twoneeff{t_{\rm w1}^{\rm eff}}
\def\twtwo{t_{\rm w2}}
\def\Upeff{\Upsilon^{\rm eff}}

\begin{document}

\runninghead{H. Takayama, H. Yoshino \& T. Komori}
{Numerical Study on Aging Dynamics in the 3D Spin-Glass Model}

\normalsize\textlineskip
\thispagestyle{empty}
\setcounter{page}{1}

\copyrightheading{Vol. 0, No. 0 (1993) 000--000}

\vspace*{0.88truein}

\fpage{1}
\centerline{\bf NUMERICAL STUDY ON AGING DYNAMICS IN THE} 
\vspace*{0.035truein}
\centerline{\bf 3D ISING SPIN-GLASS MODEL --- QUASI-EQUILIBRIUM}
\vspace*{0.035truein}
\centerline{\bf BEHAVIOUR OF SPIN AUTO-CORRELATION FUNCTIONS} 
\vspace*{0.37truein}
\centerline{\footnotesize H. TAKAYAMA, H. YOSHINO and 
T. KOMORI\footnote{Present address: Hydrographic Department, Maritime
  Safety Agency, 5-3-1 Tsukiji, Chuo-ku, Tokyo 104-0045, Japan.}} 
\vspace*{0.015truein}
\centerline{\footnotesize\it Institute for Solid State Physics, the 
University of Tokyo,}
\baselineskip=10pt
\centerline{\footnotesize\it Roppongi, Minato-ku, Tokyo 106-8666, Japan}

\vspace*{0.225truein}
\publisher{1 September 1999}

\vspace*{0.21truein}
\abstracts{Using Monte Carlo simulations, we have studied aging phenomena
in three-dimensional Gaussian Ising spin-glass model
focusing on quasi-equilibrium behavior of the spin auto-correlation
functions. Weak violation of the time translational invariance in the
quasi-equilibrium regime is analyzed in terms of effective stiffness
for droplet excitations in the presence of domain walls.
The simulated results in not only isothermal but also $T$-shift aging
processes exhibit the expected scaling behavior with respect to the
characteristic length scales associated with droplet excitations and
domain walls in spite of the fact that the growth law for these length
scales still shows a pre-asymptotic behavior compared with the
asymptotic form proposed by the droplet theory. Implications of our
simulational results are also discussed in relation to experimental
observations. 
}{}{}

\vspace*{10pt}


\vspace*{1pt}\textlineskip      
\setcounter{section}{1}
\setcounter{equation}{0}
\section{Introduction}          
\vspace*{-0.5pt}
\noindent
In recent years aging dynamics in spin glasses has been extensively 
studied.\cite{Young,VincHO,Rieger-95} A basic experimental protocol is 
isothermal aging process which is relaxation after the spin-glass
system is quenched from temperature above the spin-glass temperature
$\Tc$ to temperature $T$ below $\Tc$. One of the most interesting
results in the simulational studies on 3D short-ranged Ising
spin-glass model is that the coherence length $R_T(t)$ of the
correlation function of real two replicas at time $t$ after the quench
grows in a power-law as,\cite{Kisker-96,Marinari-98-VFDT,oursI} (but
see also\cite{Huse-91}) 
\begin{equation}
   R_T(t) \sim L_0(t/\tau_0)^{1/z}, 
\label{eq:Rt-sim}
\end{equation} 
where $L_0$ is a certain characteristic length and exponent $1/z$ 
depends on $T$. The latter was obtained as $1/z(T) \simeq 0.17T$ for
$T \le 0.7$ in our previous paper\cite{oursI} which we refer to I
hereafter. It is tempting to regard this $R_T(t)$ as a characteristic
length of the ordering process, i.e., a mean distance of domain walls
separating different pure states, which are parallel or anti-parallel
to a ground state. In this respect, it is only the droplet
theory,\cite{droplet-T,FH-88-NE} at the moment, that provides us some
concrete scaling arguments based on such characteristic length scales
as $R_T(t)$. 

According to the droplet theory by Fisher and Huse (FH),\cite{FH-88-NE} 
isothermal aging in spin glasses is associated with coarsening of 
domain walls, which is driven by successive nucleation of thermally 
activated droplets. Up to waiting time $\tw$ after the quench, domains
of averaged size $R_T(\tw)$ have grown up. Within each domain small
droplets of size $L\ (\ll R_T(\tw))$ are thermally fluctuating as they
are almost in equilibrium: a typical value of their excitation gap 
$F^{\rm typ}_L$  scales as
\begin{equation}
   F^{\rm typ}_L \sim  \Upsilon (L/L_0)^\theta,
\label{eq:FL-FH}
\end{equation}
where $\Upsilon$ is the stiffness constant, and that of free-energy 
barrier $B^{\rm typ}_{L}$ scales as 
\begin{equation}
B^{\rm typ}_L \sim \Delta (L/L_0)^\psi,
\label{eq:BL-FH}
\end{equation}
where $\Delta$ is a characteristic free-energy scale. The above 
two exponents satisfy  $\theta \le (d-1)/2$ and $\theta \le \psi 
\le d-1$ with $d$ being dimension of the system.

However, there are some droplets which touch with the domain wall so 
that their excitation gap is reduced from eq.(\ref{eq:FL-FH}). This 
effect was studied by FH by making use of some geometrical and 
probabilistic arguments. Averaged over all droplets including those 
within the domain, FH has derived a typical value of excitation gaps 
of droplets with size $L$ in the presence of the domain walls with 
mean separation $R$ as 
\begin{equation}
    F^{\rm typ}_{L,R} = \Upeff [L/R](L/L_0)^\theta,
\label{eq:domain-1}
\end{equation}
with the effective stiffness constant given by 
\begin{equation}
    \Upeff [L/R] = \Upsilon \left( 1 - c_\upsilon\left({L \over
      R}\right)^{d-\theta}\right), 
\label{eq:Upeff}
\end{equation}
where $c_\upsilon$ is a numerical constant. 

By making use of the above argument on the effective stiffness 
constant, the spin auto-correlation function 
\begin{equation}
   C_T(\tau; \tw) = \overline{ C_{i,T}(\tau; \tw)} \ \ \ {\rm with} \ \ \ 
C_{i,T}(\tau; \tw) = \langle S_i(\tau+\tw)S_i(\tw)\rangle_T,
\label{eq:def-Cor}
\end{equation}
is evaluated as the following. Some further details of its derivation
are described in our separated paper\cite{oursII} which we refer to II 
hereafter. A droplet of size $L$ enclosing spin $S_i$ contributes to
$C_{i,T}(\tau; \tw)$ as  
\begin{equation} C_{i,T}(\tau; \tw) \simeq \left\{
     \begin{array}{ll}
        1  & {\rm for} \ \ \tau_L(i) > \tau, \\
        <S_i>_T^2 \ \simeq \ 1 - 4{\rm exp}(-F_{L,R}(i)/T) \ \ & 
{\rm for} \ \ \tau_L(i) < \tau, 
\nonumber
     \end{array}\right.
\label{eq:Cori-cal}
\end{equation}
where $\tau_L(i)=\tau_0{\rm exp}(B_L(i)/T)$ is the relaxation time 
of the droplet whose free-energy barrier is $B_L(i)$. The averages 
(denoted by the over-line in eq.(\ref{eq:def-Cor})) over site $i$ and 
different realizations of interactions (samples) are all taken into 
account by the probability distribution of free-energy gaps 
$F_{L,R}(i)$ which is assumed to scale as, 
\begin{equation}
\rho_{L,R}(F) \simeq {1 \over F^{\rm
  typ}_{L,R}}{\tilde \rho} \left( { F \over F^{\rm typ}_{L,R} } \right), 
\label{eq:rho-FH}
\end{equation}
where a scaling function ${\tilde \rho}(x)$ satisfies ${\tilde \rho}(0) 
> 0$. Then, gathering the multiplicative contributions of droplets
with various sizes, $L=2^nL_0; n=0,1,2,...$, enclosing spin $S_i$, we
obtain
\begin{eqnarray}
    1 - C_T(\tau; \tw)  & \sim & 
  \int_{L_0}^{L_T(\tau)} {{\rm d}L \over L} 
  \frac{{\tilde \rho}(0) T}{(L/L_0)^\theta}
  \frac{1}{\Upeff[L/R_T(\tw)]}\\
\label{eq:cal-Cor-3}
   & \sim & 1 - C_{{\rm eq},T}(\tau) +  
\frac{c_1\tilde{\rho}(0)T}{\Upsilon(L_T(\tau)/L_0)^\theta}
\left( \frac{L_T(\tau)}{R_T(\tw)} \right)^{d-\theta}+\dots, 
\label{eq:cal-Cor-5}
\end{eqnarray}
where $c_1$ is a numerical constant and
\begin{equation}
1-C_{{\rm eq},T}(\tau) \sim 
\frac{T{\tilde \rho}(0)}{\Upsilon}\int_{L_{0}}^{L_T(\tau)} 
\frac{dL}{L}\frac{1}{(L/L_0)^\theta}
=\frac{T{\tilde \rho}(0)}{\theta\Upsilon}\left[ 1 - \left( {L_0 \over
  L_T(\tau)} \right)^\theta \right].  
\label{eq:Coreq-cal}
\end{equation}
Here $R_T(t)$ is given by
\begin{equation}
   R_T(t) \sim L_0[(T/\Delta){\rm ln}(t/\tau_0)]^{1/\psi}. 
\label{eq:Rt-FH}
\end{equation}
and $L_T(\tau)$ by the same expression with $t$ replaced by $\tau$ 
in accordance with eq.(\ref{eq:BL-FH}). The latter is the 
characteristic length of droplets which can fluctuate within time 
scale $\tau$ at temperature $T$. The above scaling expression 
eq.(\ref{eq:cal-Cor-5}) is appropriate in the regime 
$L_T(\tau) \ll R_T(\tw)$ which defines the quasi-equilibrium regime 
of present interest. The third term in eq.(\ref{eq:cal-Cor-5})
represents the leading correction to equilibrium behaviour which
weakly violates the time translational invariance in this regime.  

The simulational result eq.(\ref{eq:Rt-sim}) differs from 
eq.(\ref{eq:Rt-FH}). A possible interpretation of this discrepancy
which we will pursue in the present work is as follow. The difference
in the growth laws of $R_T(t)$ and $L_T(\tau)$ is simply due to two
different time windows which one observes the aging processes; one is  
an asymptotic regime for which the droplet theory is constructed, and
the other a pre-asymptotic regime that the simulations can observe.
However, we consider that the scaling expressions written in terms of 
$R_T(t)$ and $L_T(\tau)$, such as eq.(\ref{eq:cal-Cor-5}), are common
to both regimes. A similar interpretation has been already proposed
for some aspects of aging dynamics.\cite{Huse-91,Rieger-93}  In our
paper II we have shown by detailed analysis on the spin
auto-correlation function that the droplet picture in the sense
described above in fact holds in the quasi-equilibrium of the
isothermal aging in the 3D short-ranged Ising spin-glass model. 

The main purpose of the present work is to demonstrate that the same
picture consistently describes behavior of the aging dynamics in the
{\it quasi-equilibrium regime} of the so-called $T$-shift
process.\cite{VincHO,Nordblad} The latter is another typical
experimental protocol of aging: first to let the system age
isothermally at $T_1$, to change temperature to $T_2$ at waiting time
$\twone$, to let the system age isothermally at $T_2$ in a period
$\twtwo$ at which a small probing field is added, and then to measure 
response of the system at time $\tau$ afterwards. We may write the
corresponding correlation function as $C_{T_2, T_1}(\tau; \twtwo;
\twone)$. 

One of most interesting problems in the $T$-shift process is whether
or how we can observe chaotic nature of the spin-glass phase which is
derived from the droplet theory;\cite{droplet-T,BrayM-chaos} namely, 
there exists such a characteristic length scale $L_{|T_1-T_2|}$,
called the overlap length, that the equilibrium spin-glass orders at
$T_1$ and $T_2$ differ entirely from each other in the length-scale
larger than $L_{|T_1-T_2|}$. In the present work, as for the first
step of our study on this problem, we have analyzed  
$C_{T_2, T_1}(\tau; \twtwo; \twone)$ with $\twtwo=0$ in the {\it
quasi-equilibrium regime} which is specified by $\tau \ll \twone$. Up
to now even a precursor of the chaotic nature has been ascertained
within the time scale of our simulation.  Instead, as mentioned above,
the simulated results are consistently interpreted by the droplet
picture without introduction of $L_{|T_1-T_2|}$. In other words,
$L_{|T_1-T_2|}$, if it exists, is much larger than the length scale of
our simulation.
    
After describing our model and numerical method in the next section,
we briefly review the results on behaviors of the correlation function 
in the quasi-equilibrium regime of the isothermal aging in \S 3. We
then present and discuss the corresponding results in the $T$-shift
aging process in \S4. In the last section we argue implications of our 
simulational results on the experimental observations based on the
fluctuation-dissipation theorem.  

\setcounter{section}{2}
\setcounter{equation}{0}
\section{Model and Method}

We have carried out standard heat-bath Monte Carlo simulation on 
aging phenomena in the 3D Ising spin-glass model with Gaussian
nearest-neighbor interactions with zero mean and variance $J=1$.
The spin-glass transition temperature is numerically determined 
most recently as $\Tc=0.95\pm 0.04$.\cite{Marinari-cd98-PS}
The data we will discuss below are obtained at $T=0.5 \sim 0.8$ in 
$L_{\rm s}=24$ system averaged over 160 samples with one MC run for
each sample. In our previous work,\cite{oursI} hereafter referred to
I, it was confirmed that finite-size effects do not appear within our
time window ($\nle 2 \times 10^{5}$ MCS) for these parameters. Also in
I we showed from the finite-size-scaling analysis of relaxation of the 
energy per spin in isothermal aging that the exponent $\theta$ in 
eq.(\ref{eq:FL-FH}) at $T=0.7, 0.8$ is estimated as 
$\theta = 0.20 \pm 0.03$ which coincides with the result of the defect 
energy analysis at $T=0$.\cite{BM-84-theta}

\setcounter{section}{3}
\setcounter{equation}{0}
\section{Results on Isothermal Aging}

It is now well established that the correlation function 
$C_T(\tau; \tw)$ in isothermal aging consists of two characteristic
time regimes, one is the range $\tau \ll \tw$ which is the 
quasi-equilibrium regime mentioned in \S 1, and the other 
$\tau \gg \tw$ called as the out-of-equilibrium (or aging) regime. 
Our interest here is in its behaviors in the former regime. As shown
in our paper II, characteristic features of $C_T(\tau; \tw)$ in the
quasi-equilibrium regime are more easily understood when $\DelCone{}$
are analyzed as a function of $\tw/\tau$ with $\tau$ being fixed. It
turns out that such data points of different $\tau$ lie top on each
others when they are shifted vertically by proper amounts
$\alpha_T(\tau)$. The results of such a one-parameter scaling are
demonstrated in Fig.~1 where we plot  
\begin{equation}
  \DelC{} = \DelCone{} - \alpha_T(\tau) 
\label{eq:DelC}
\end{equation}
against $\tw/\tau$. The $\{ \alpha_T(\tau) \}$ thus determined are 
presented in the inset of Fig.~1 where we also show the results of the
same analysis at different $T$'s. They appear as almost linear
functions of ln$\tau$. 

In order to compare the above results with the scaling expressions 
eqs.(\ref{eq:cal-Cor-5}) and (\ref{eq:Coreq-cal}), we only need to
notice the fact that factor $(L_{0}/L_T(\tau))^{\theta}$ can be
practically expanded as 
\begin{equation}
\left( {L_0 \over L_T(\tau)} \right)^\theta 
= \left( {\tau_0 \over \tau} \right)^{\theta/z(T)}
\simeq 1 - { \theta \over z(T) }{\rm ln}\left( {\tau \over \tau_0}
\right), 
\label{eq:expand}
\end{equation}
since $\theta/z(T)$ is small as compared with $1/{\rm
ln}(\tau_{\rm max}/\tau_0)$, where $\tau_{\rm max}\ (=512)$ is the 
maximum $\tau$ in the present observation where we put $\tau_0=1$. 
Indeed, $\theta/z(T) \simeq 0.02$ with $\theta = 0.20 \pm 0.03$ and
$1/z(T=0.6)\simeq 0.102$, while $1/{\rm ln}(\tau_{\rm max}/\tau_0) 
\simeq 0.160$. This factor in eq.(\ref{eq:cal-Cor-5}) can be
approximated by its leading term ($=1$), so that the last term in
eq.(\ref{eq:cal-Cor-5}) becomes a function of only
$L_T(\tau)/R_T(\tw)$, which is given by 
\begin{equation}
\Delta C_T(\tau; \tw) \propto (L_T(\tau)/R_T(\tw))^\kappa, 
\label{eq:kappa}
\end{equation}
with $\kappa = d - \theta$. This is consistent with the fact that the 
one-parameter scaling shown in Fig.~1 does work well. The leading term 
of eq.(\ref{eq:Coreq-cal}), on the other hand, comes out from the
second term in eq.(\ref{eq:expand}), and we obtain 
\begin{equation}
\alpha_T(\tau) = 1-C_{{\rm eq},T}(\tau) 
\simeq  \frac{T{\tilde \rho}(0)}{\Upsilon}\frac{1}{z(T)}
\ln \left(\frac{\tau}{\tau_{0}} \right).
\label{eq:alpha-cal}
\end{equation}
This is quite consistent with nearly linear increase of 
$\alpha_T(\tau)$ shown in the inset of Fig.~1. Furthermore we obtain 
$\partial \alpha_T(\tau)/\partial {\rm ln}\tau \propto T^{2.4}$ from 
the data at four temperatures shown in the figure. This
is compatible with eq.(\ref{eq:alpha-cal}) which predicts the
corresponding slope proportional to $T^2$ (note that $1/z(T) \propto T$). 
The extra $T$-dependence, which yields the power $T^{2.4}$ larger than 
$T^2$, may be attributed to that of the factor ${\tilde
  \rho}(0)/\Upsilon$. 

\vspace*{13pt}
\begin{center}
\leavevmode\epsfxsize=85mm
\epsfbox{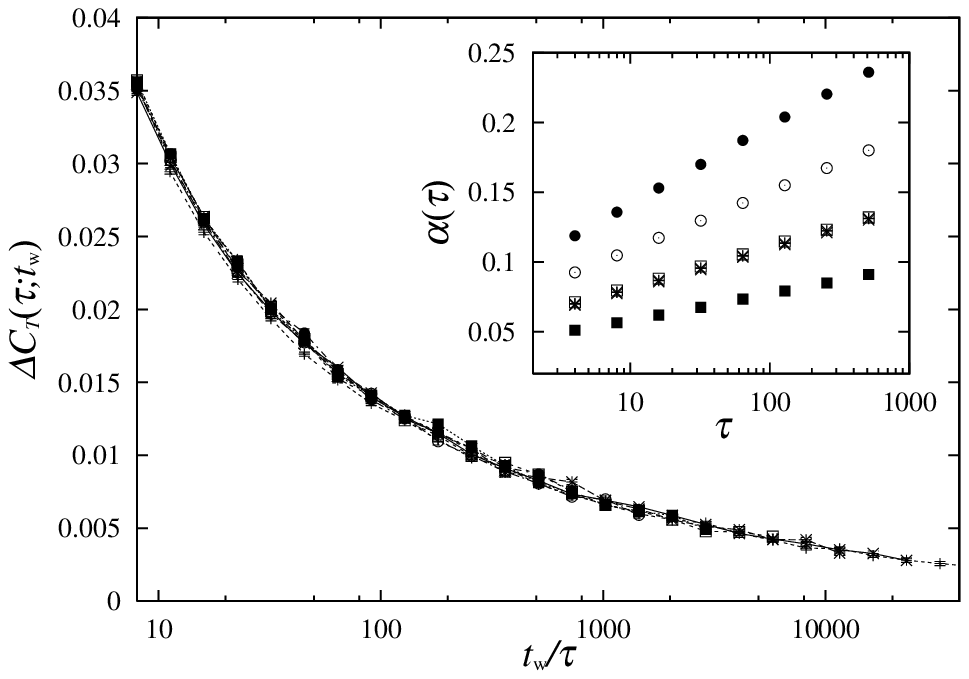}
\end{center}
\vspace*{10pt}
\fcaption{
The function $\DelC{}$ as a function of $\tw/\tau$ obtained 
in isothermal aging at $T=0.6$. In the inset we show 
$\{ \alpha(\tau) \}$ thus extracted at $T=0.5, 0.6, 0.7$ and $0.8$ 
from bottom to top. Four symbols at $T=0.6$ represents four sets of 
$\{ \alpha(\tau) \}$ which yield significantly different values of 
$\kappa$ in eq.(\ref{eq:kappa}).
}
\vspace*{13pt}

As for the determination of value $\kappa$ in eq.(\ref{eq:kappa}),
however, a more careful treatment is required since our simulational 
time window is not enough wide to extract precisely the limiting 
behavior of $\DelC{}$ at $L_T(\tau)/R_T(\tw) \rightarrow 0$. By the
detailed analysis described in II, the values of $\kappa$ acceptable
within accuracy of the present simulated data are estimated as $\kappa
\simeq 2.3 \sim 3.1$. The result is compatible with the expected value
$d - \theta \simeq 2.8$, though the data of $\DelC{}$ alone cannot
specify the value of $\theta$. Let us remark also that this ambiguity
in $\kappa$ little affects the values of $\{ \alpha_T(\tau) \}$, in
particular, their slopes with respect to ln$\tau$. This is
demonstrated in the inset of Fig.~1 by drawing four sets of 
$\{ \alpha_T(\tau) \}$ at $T=0.6$ which yield $\kappa \simeq 2.2, 2.6,
3.0,$ and $3.3$, respectively. These circumstances are similar at
other temperatures we have examined. Therefore we show the
representative results of $\{ \alpha_T(\tau) \}$ in the inset of
Fig.~1 which give rise to $\kappa \simeq 2.8 \sim 3.0$.

So far we have demonstrated that our simulated data of $C_T(\tau; \tw)$ 
are consistently interpreted by the scaling ansatz 
eq.(\ref{eq:cal-Cor-5}), when the growth law of $R_T(\tw)$ 
eq.(\ref{eq:Rt-FH}) in the droplet theory is replaced by 
eq.(\ref{eq:Rt-sim}). We regard that the former describes the growth
law in the asymptotic regime closer to equilibrium, while the latter
does in the pre-asymptotic regime, and that the common scaling forms
such as eq.(\ref{eq:cal-Cor-5}) hold in both regimes. 

\setcounter{section}{4}
\setcounter{equation}{0}
\section{Results on $T$-Shift Process}

Among the $T$-shift aging processes described in \S 1, here we restrict 
ourselves to analysis on $C_{T_2, T_1}(\tau; \twtwo; \twone)$ with
$\twtwo=0$, which we abbreviate as $C_{T_2, T_1}(\tau; \twone)$. At
time $\twone$ domains with mean separation $R_{T_1}(\twone)$ have
grown up. It has been confirmed\cite{Komori} that, by the $T$-shift to
$T_2$ at $\twone$, the correlation length of the replica-overlap
function does not decrease. Rather it continues to increase though
relatively gradually (rapidly) when $T_1 >(<)\ T_2$, and finally it
merges to the isothermal curve $R_{T_2}(t)$ at $T_2$ at around
$\tau=\twoneeff$ which we define by the condition
\begin{equation}
 R_{T_2}(\twoneeff) \simeq R_{T_1}(\twone),\ \ \ {\rm or}\ \ \ 
 \twoneeff = a\twone^{z(T_2)/z(T_1)},
\label{eq:tweff}
\end{equation} 
with $a$ being a numerical constant rather close to unity. This means
that $R_{T_1}(\twone)$ with $\twone$ we have examined does not exceed
the overlap length $L_{|T_1-T_2|}$ mentioned in \S 1. That
$L_{|T_1-T_2|}$, in case it exists, is much larger than the length
scale of the presently available simulation agrees with the previous
work.\cite{Kisker-96}  In the following, therefore, we analyze our data 
discarding the chaotic effect.

Then from the observed results mentioned above, we can simply think of
a scenario that, in the time range $\tau \ll \twoneeff$ after the 
$T$-shift, there exist domains nearly in equilibrium at $T_1$ and 
with mean separation of $R_{T_1}(\twone)$, and droplets up to mean 
size of $L_{T_2}(\tau)$ are fluctuating by thermal noises of 
temperature $T_2$. In this situation with $R_{T_1}(\twone) \gg 
L_{T_2}(\tau)$, which we call the {\it quasi-equilibrium regime} in 
the present $T$-shift aging process, $R_{T_1}(\twone)$ plays a role 
of $R_T(\tw)$ in equations which correspond to
eqs.(\ref{eq:cal-Cor-3}), (\ref{eq:cal-Cor-5}) and
(\ref{eq:kappa}). More explicitly, we expect that, if 
$1 - C_{T_2, T_1}(\tau; \twone)$ for various $\tau$ are plotted
against $\twoneeff/\tau$, they look quite similar to the
$1-C_T(\tau;\tw)$  versus $\tw/\tau$. Furthermore, as is the case for
the latter shown in Fig.~1, $1 - C_{T_2, T_1}(\tau; \twone)$ are
expected to lie on a universal curve by vertical shift,
$\alpha_{T_2,T_1}(\tau)$, of the data for each $\tau$. Indeed this is
the case  as shown in Fig.~2 where we plot 
\begin{equation}
  \DelCtwo{} = \DelConetwo{} - \alpha_{T_2,T_1}(\tau) 
\label{eq:DelCtwo}
\end{equation}
against $\twoneeff/\tau$.

\vspace*{13pt}
\begin{center}
\leavevmode\epsfxsize=85mm
\epsfbox{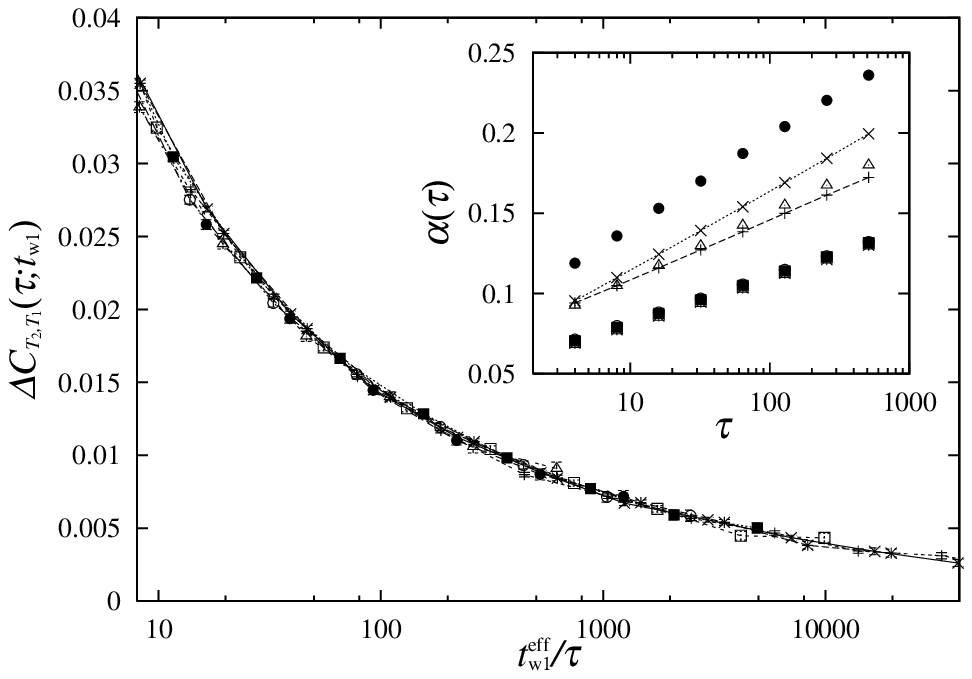}
\end{center}
\vspace*{10pt}
\fcaption{The plot $\DelCtwo{}$ versus 
$\twoneeff/\tau$ for the $T$-shift process with $T_1=0.8$ and 
$T_2=0.6$. The functions $\alpha_{T_2,T_1}(\tau)$ thus extracted 
are presented in the inset: symbols with the dotted line represent 
$\alpha_{T_2=0.6,T_1=0.8}(\tau)$ and those with 
the broken line $\alpha_{T_2=0.8,T_1=0.6}(\tau)$, while only the 
symbols do the isothermal $\alpha_T(\tau)$ with $T=0.6, 0.7$ and 
$0.8$ from bottom to top. 
}
\vspace*{13pt}

 The vertical shifts $\alpha_{T_2,T_1}(\tau)$ in eq.(\ref{eq:DelCtwo})
depend on both $T_1$ and $T_2$. To explain this let us go back to an 
equation for the local correlation function, which corresponds to
eq.(\ref{eq:Cori-cal}) for the isothermal process. In the present
$T$-shift process deviation of the function from unity occurs also
when a droplet enclosing spin $S_i$ has relaxation time $\tau_L(i)$
which is smaller than $\tau$, but the value it relaxes to is now given
by  
\begin{equation}
 C_{i; T_1,T_2}(\tau; \twone)   \simeq <S_i>_{T_1}<S_i>_{T_2}
\simeq  1 - 2[{\rm exp}(-F_{L,R}(i)/T_1) + {\rm exp}(-F_{L,R}(i)/T_2)], 
\label{eq:cal-Cori2}
\end{equation}
for $\tau > \tau_L(i)$, since $S_i$ here is regarded in equilibrium of 
temperatures $T_1$ and $T_2$ at times $\tau=0$ and $\tau >\tau_L(i)$,
respectively. Therefore we obtain, instead of eq.(\ref{eq:alpha-cal}),
\begin{equation}
\alpha_{T_2,T_1}(\tau)
\sim \frac{1}{2} \left( 
\frac{T_1{\tilde \rho}(0)}{\Upsilon} + 
\frac{T_2{\tilde \rho}(0)}{\Upsilon} \right) 
\frac{1}{z(T_2)}
\ln \left(\frac{\tau}{\tau_{0}} \right).
\label{eq:alpha2-cal}
\end{equation}
This expression, combined with eq.(\ref{eq:alpha-cal}), consistently
explains the relative orders of slopes $\partial \alpha(\tau)/
\partial {\rm ln}\tau \ (=\Delta\alpha)$ of various $\alpha(\tau)$
shown in the inset of Fig.~2; namely, $\Delta\alpha_{0.8} >
\Delta\alpha_{0.8,0.6} > \Delta\alpha_{0.7} > \Delta\alpha_{0.6,0.8} >
\Delta\alpha_{0.6}$. 

\vspace*{13pt}
\begin{center}
\leavevmode\epsfxsize=75mm
\epsfbox{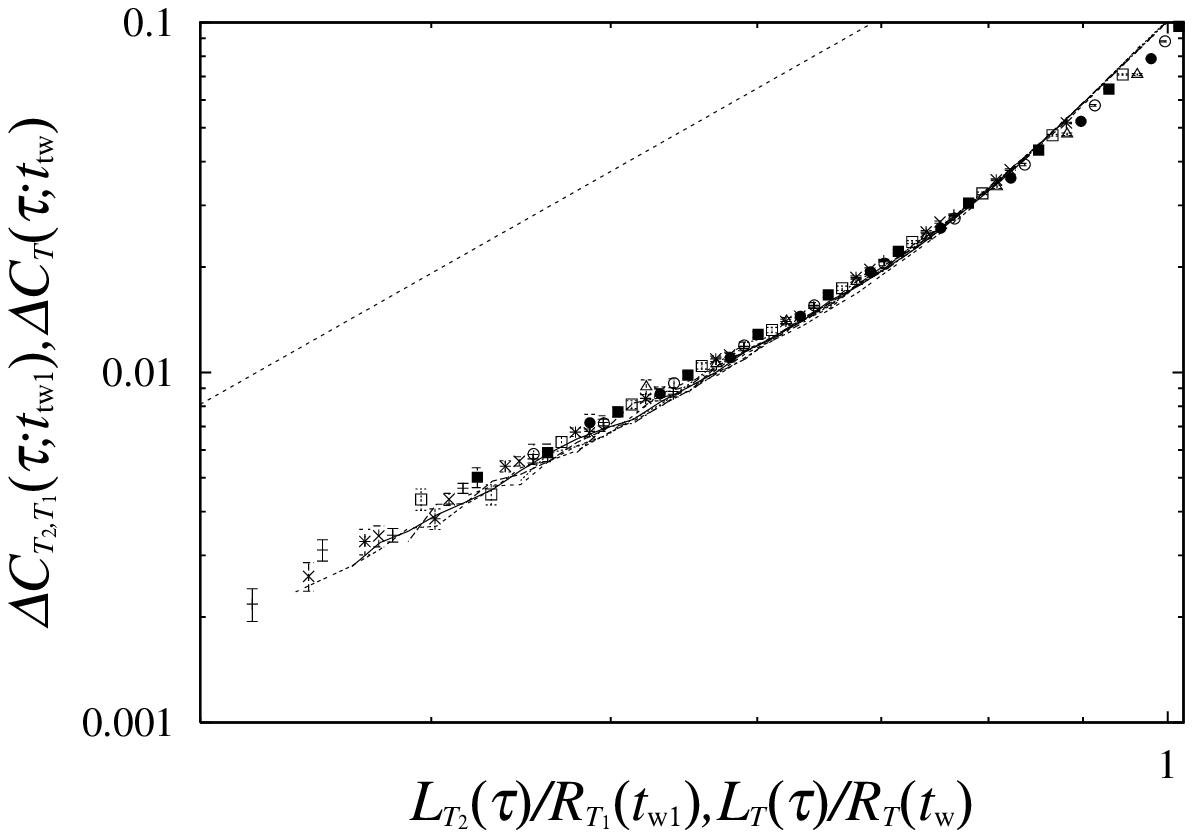}
\end{center}
\vspace*{10pt}
\fcaption{The double logarithmic plot of $\DelCtwo{}$ versus 
$L_{T_2}(\tau)/R_{T_1}(\twone)$ of the $T$-shift process for different 
$\tau$'s are plotted by the different symbols (with $T_1=0.8$ and 
$T_2=0.6$). For comparison $\DelC{}$ versus $L_{T}(\tau)/R_{T}(\tw)$ 
of the isothermal aging for different $\tau$'s are plotted by the 
different curves (at $T=0.6$). The slope of the dotted line is $3.0$. 
}
\vspace*{13pt}

Finally we show in Fig.~3 the double logarithmic plot of $\DelCtwo{}$ 
versus $(\tau/\twoneeff)^{1/z(T_2)}$ which is proportional to 
$L_{T_2}(\tau)/R_{T_1}(\twone)$. For comparison, we also plot $\DelC{}$ 
versus $(\tau/\tw)^{1/z(T)} \ (\propto L_{T}(\tau)/R_{T}(\tw)$) of the 
isothermal aging. Similarly to the latter case, there remains a 
relatively large ambiguity in the value of the slopes, i.e., the 
exponent which corresponds to $\kappa$ in eq.(\ref{eq:kappa}). 
But the result is also compatible with $\kappa = d - \theta$ within 
accuracy of the present simulations. 

\setcounter{section}{5}
\setcounter{equation}{0}
\section{Discussions}

Our simulational results on the correlation function $C(t,t') \ 
(=C_T(\tau; \tw)$ with $t=\tau+\tw$ and $t'=\tw$) so far described 
are closely related to the response function $G(t,t')$ via the 
fluctuation-dissipation theorem (FDT) 
\begin{equation}
G(t,t') = {1 \over T}{\partial C(t,t') \over \partial t'}, 
\label{eq:FDT-1}
\end{equation}
and so the magnetic susceptibilities frequently measured in 
experiments. That the above FDT holds at least approximately in the 
quasi-equilibrium regime even with weak violation of the 
time-translational invariance (TTI) has been carefully checked 
experimentally.\cite{AlbaHOR-87} It has been also demonstrated 
in the simulational work by Franz and Rieger,\cite{FranzR} though 
the authors did not mention it explicitly: their data which exhibit 
the FDT violate the TTI, i.e., depend also on $\tw$. 
 
From eq.(\ref{eq:FDT-1}) the following relation between the 
out-of-phase component of the ac susceptibility 
$\chi''_T(\omega; \tw)$ and $C_T(\tau_\omega; \tw)$ in the 
isothermal aging is derived,
\begin{equation}
     \chi''_T(\omega; \tw) \simeq - \left.{\pi \over 2T}
    {\partial \over \partial {\rm ln} \tau}
     C_T(\tau;\tw)\right|_{\tau = \tau_\omega},
\label{eq:chi-C-1}
\end{equation}
where $\tau_\omega = 2\pi/\omega$. Its derivation and implication on 
the experimental results, such as the $\omega\tw$-scaling of 
$\chi"_T(\omega; \tw)$, are described in II. 

Here let us emphasize that from eq.(\ref{eq:FDT-1}) we obtain 
also the relation between $C(\tau; \tw)$ and the zero-field cooled 
susceptibility $\chi^{\rm ZFC}(\tau;\tw)$ as
\begin{equation}
\chi^{\rm ZFC} (\tau; \tw) = \int_{\tw}^{\tau + \tw}{\rm d}t'G(\tau+\tw, t') 
= {1 \over T}[1 - C(\tau; \tw)]. 
\label{eq:FDT-chi}
\end{equation}
Thus $1 - C_T(\tau; \tw)$ in the isothermal aging, which we have 
analyzed in \S 3, is directly related to $\chi^{\rm ZFC}_T(\tau;\tw)$ 
in the quasi-equilibrium regime. The similar relation is expected to 
hold in the {\it quasi-equilibrium regime} of the $T$-shift process 
which we have discussed in \S 4, i.e., when the response is examined 
by adding a probing field from $\twtwo=0$. Although 
$\chi^{\rm ZFC}_T(\tau; \tw)$ has been measured frequently since the 
pioneering work by Lundgren {\it et al}\cite{Lundgren} and 
$\chi^{\rm ZFC}_{T_2,T_1}(\tau; \tw)$ in the $T$-shift process with
$\twtwo=0$ has been also examined,\cite{Nordblad} most of the
measurements have been focused on its crossover behaviour between the
quasi-equilibrium and aging regimes. We believe that the detailed 
analyses on these $\chi^{\rm ZFC}(\tau; \tw)$ are of quite importance 
since they will present us evidences for (or against) the droplet 
picture we have discussed in the present work.

To conclude, we have simulated aging dynamics in the 3D Ising 
spin-glass model, and analyzed behaviors of the spin auto-correlation 
function in the quasi-equilibrium regime of the $T$-shift aging
processes. The simulated results, which are considered to represent
aging properties in the pre-asymptotic regime, do not exhibit even a
precursor of the chaotic nature of the spin-glass phase. Instead, they 
are satisfactorily explained by the scaling argument based on the
characteristic length scales $R_{T_1}(\twone)$ and $L_{T_2}(\tau)$,
i.e., that of domains grown up till $\twone$ at $T_1$ and that of
droplet excitations in time scales of $\tau$ at $T_2$. However their
growth law itself differs from the prediction by the droplet theory
which is constructed for the asymptotic regime. We have conjectured
that the experimental observation on $\chi^{\rm ZFC}(\tau; \tw)$ in
the quasi-equilibrium regime is of quite interest to check this
droplet picture.  

\nonumsection{Acknowledgment}
Two of the present authors (T. K. and H. Y.) were supported by
Fellowships of Japan Society for the Promotion of Science for Japanese
Junior Scientists. This work is supported by a Grant-in-Aid for
International Scientific Research Program
(\#10044064) and by a Grant-in-Aid
for Scientific Research Program (\#10640362), from the Ministry of
Education, Science and Culture. Most of the present simulation has
been done on RANDOM at Material Design and Characterization
Laboratory, Institute for Solid State Physics, University of Tokyo. 

\nonumsection{References}

\end{document}